%% file: main.tex
\documentclass[
  reprint,
  prl,
  amsmath, amssymb,
  aps,
  superscriptaddress,
  longbibliography
]{revtex4-2}

\usepackage{graphicx}
\usepackage{dcolumn}
\usepackage{bm}

\usepackage{ragged2e}
\usepackage{etoolbox}
\usepackage{comment}
\usepackage{multirow}
\usepackage{textcomp}
\usepackage{quantikz}

\usepackage{hyperref}

\begin{document}

\preprint{APS/123-QED}

\title{A scalable quantum-neural hybrid variational algorithm for ground state estimation}

\author{Minwoo Kim}
 \affiliation{Dept. of Computer Science and Engineering, Seoul National University, Seoul 08826, South Korea}
 \affiliation{NextQuantum, Seoul National University, Seoul 08826, South Korea}
\affiliation{Automation and Systems Research Institute, Seoul National University, Seoul 08826, South Korea}

\author{Kyoung Keun Park}%
\affiliation{Dept. of Computer Science and Engineering, Seoul National University, Seoul 08826, South Korea}
\affiliation{NextQuantum, Seoul National University, Seoul 08826, South Korea}
\affiliation{Automation and Systems Research Institute, Seoul National University, Seoul 08826, South Korea}

\author{Uihwan Jeong}%
\affiliation{Dept. of Computer Science and Engineering, Seoul National University, Seoul 08826, South Korea}
\affiliation{NextQuantum, Seoul National University, Seoul 08826, South Korea}
\affiliation{Automation and Systems Research Institute, Seoul National University, Seoul 08826, South Korea}

\author{Sangyeon Lee}%
\affiliation{NextQuantum, Seoul National University, Seoul 08826, South Korea}
\affiliation{Dept. of Electrical and Computer Engineering, Seoul National University, Seoul 08826, South Korea}

\author{Taehyun Kim}
\email{taehyun@snu.ac.kr}
\affiliation{Dept. of Computer Science and Engineering, Seoul National University, Seoul 08826, South Korea}
\affiliation{NextQuantum, Seoul National University, Seoul 08826, South Korea}
\affiliation{Automation and Systems Research Institute, Seoul National University, Seoul 08826, South Korea}
\affiliation{Institute of Computer Technology, Seoul National University, Seoul 08826, South Korea}
\affiliation{Institute of Applied Physics, Seoul National University, Seoul 08826, South Korea}

\date{\today}


\begin{abstract}
We propose the unitary variational quantum-neural hybrid eigensolver (U-VQNHE), which improves upon the original VQNHE by enforcing unitary neural transformations. The non-unitary nature of VQNHE causes normalization issues and divergence of the loss function during training, leading to exponential scaling of measurement overhead with qubit number. U-VQNHE resolves these issues, significantly reduces required measurements, and retains improved accuracy and stability over standard variational quantum eigensolvers.
\end{abstract}

\maketitle

\paragraph*{Introduction}

Variational quantum eigensolver (VQE)~\cite{peruzzo14, tilly2022} is a key development in the field of quantum algorithms. VQE addresses one of the most challenging problems in quantum chemistry—calculating the ground state energy of molecular systems~\cite{mcardle20, lim24, AlBalushi24}. It employs a hybrid approach, using quantum computers to provide a parametrized quantum circuit (PQC)~\cite{du22, benedetti19} for evaluating expectation values, while classical computers optimize the circuit parameters. Significance of VQE lies in its scalability and adaptability to noisy intermediate-scale quantum (NISQ) hardware~\cite{preskill18}, making it more hardware-friendly, especially in terms of circuit depths, compared to other ground-state estimating algorithms based on quantum phase estimation~\cite{aspuru05, wang24}.

When evaluating ground state energy, a major consideration is designing an ansatz that efficiently expresses the desired solution while maintaining reasonable computational cost. Researchers have sought to construct ansatze that express the physical nature of systems, which are so-called physics-inspired ansatze, achieving excellent accuracies~\cite{mccaskey19, anand22, zhang22}. However, these often require significant depth and long-range connectivity, which quantum computers in NISQ era struggle with.

To address practical limitations, researchers developed the hardware-efficient ansatze. This approach is designed for implementation using simple native gates supported by quantum hardware, considering limited connectivity to avoid long-range entangling gates. While more easily implementable on NISQ devices, they yield less accurate values than the physics-motivated ansatze~\cite{kandala17, leone22, Casanova24}.

Although their expressiveness is limited, it has been shown that classical post-processing can enhance their capabilities~\cite{zhang22, gunlycke2024, mazzola2019, benfenati2021}. A novel approach, the variational quantum-neural hybrid eigensolver (VQNHE)~\cite{zhang22}, exploits a neural network to apply additional transformations to the quantum state. The neural network processes each quantum circuit measurement as a binary string, producing a single-numbered output without exponential computational overhead. The expectation value of the Hamiltonian is then calculated by combining measurement results with neural network outputs. This method, using a statevector simulator, has demonstrated significant improvements for transverse-field Ising model (TFIM)~\cite{pfeuty70} and molecular Hamiltonian simulations compared to the original VQE using hardware-efficient ansatze.

Despite improved performance achieved with VQNHE, this combination of quantum and classical processes has a critical defect. We have discovered that it can optimize the neural network to yield a very large negative expectation value of the given Hamiltonian regardless of the actual ground state energy, and that such a phenomenon can only be prevented with an exponential number of quantum circuit measurements, causing scalability bottlenecks in terms of practical implementation. To address the challenge at hand, we propose a novel way of utilizing a neural network to perform a nontrivial unitary transformation in a complex domain. This structure circumvents the need for normalization, which has been the primary reason for the critical divergence during neural network training. With this advanced method, we eliminate the need for an exponentially large number of circuit shots, thereby improving its computational efficiency.

Furthermore, even when a sufficient number of measurements is provided to avoid divergence, we observe that VQNHE can still converge to values that significantly deviate from the exact ground state energy. Our results demonstrate that, due to the tightly constrained norms imposed on the neural network outputs, our algorithm exhibits enhanced stability and reduced deviation from the exact ground state value. This enhancement enables end-to-end scalability of our method with respect to both computational resources and the number of required measurements.

\paragraph*{Summary of the VQNHE.}

\begin{figure}
\centering
\includegraphics[width=0.75\linewidth]{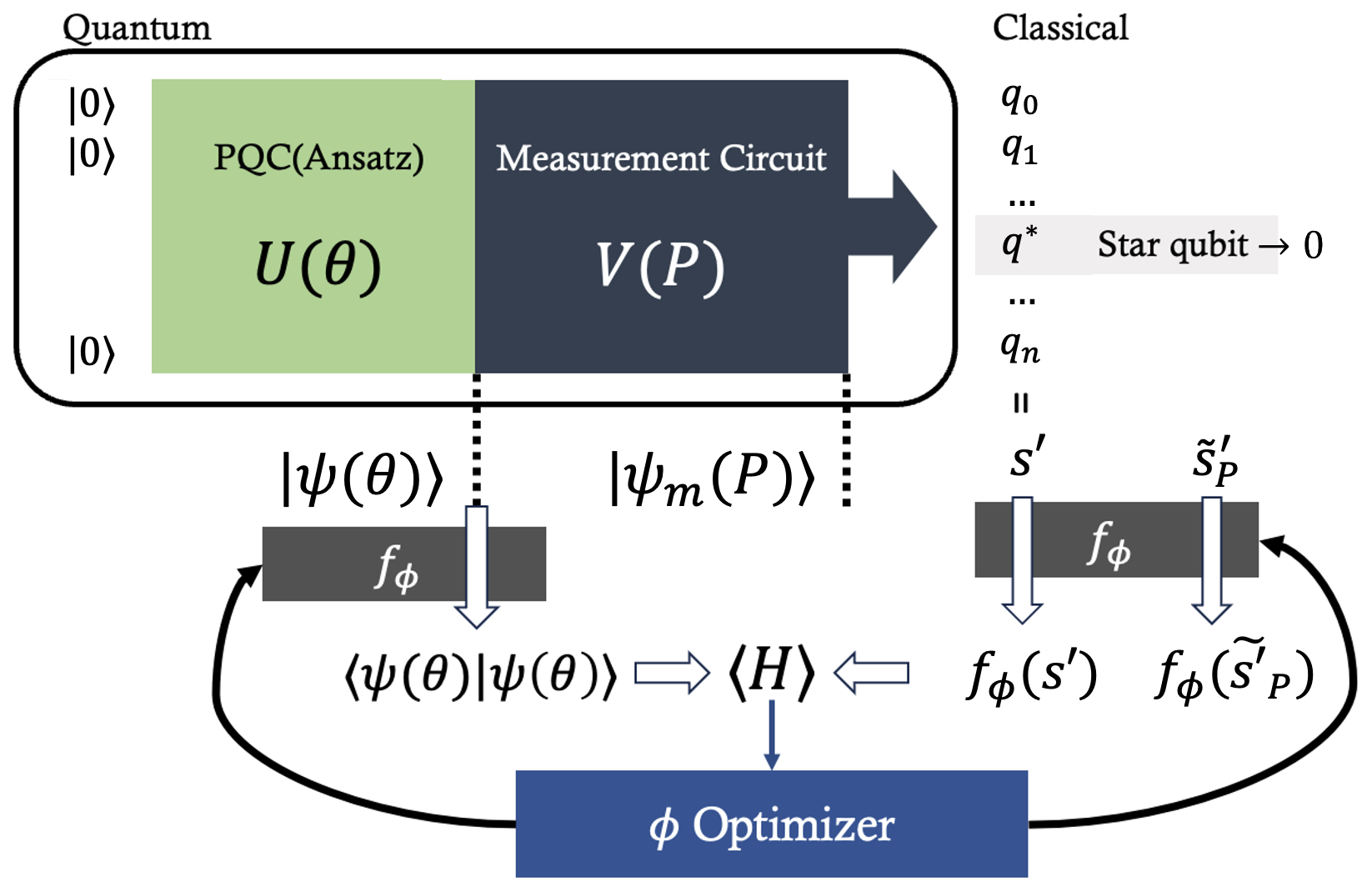}
\caption{Flowchart of the VQNHE algorithm. The round box indicates the quantum part of the algorithm. The parameters $\theta$ of the quantum circuit are trained before the parameters $\phi$ of the neural network, whose process is not shown in the figure.}
\label{vqnhe}
\end{figure}

VQE aims to compute the ground state energy of a given Hamiltonian. It consists of two main parts: quantum circuit that implements parametrized ansatz and classical optimizer that trains the parameters to yield the lowest possible eigenvalue~\cite{Bausch23, Meyer22}. One can express $n$-qubit parametrized ansatz circuit with parameters $\theta$ as a unitary operation $U(\theta)$ that acts on the initial state $|0\rangle^{\otimes n}$ to generate $|\psi(\theta)\rangle = U(\theta)|0\rangle^{\otimes n}$. The goal is to approximate the ground state energy of a Hamiltonian written in a form
\begin{equation}
    \hat{H} = \sum_{P \in \{I, X, Y, Z\} ^ {\otimes n}} h_P \hat{P}, h_P \in \mathbb{R}.
\end{equation}
For each Pauli term composing the Hamiltonian, its expectation value is evaluated by measuring the ansatz circuit in the corresponding basis. After measuring the expectation value, one uses a classical optimizer~\cite{Bausch23, Meyer22} or quantum gradients~\cite{wierichs22} to find the parameter set that makes the ansatz an optimal solution for the Hamiltonian.

The VQNHE algorithm appends a neural network to VQE by feeding the quantum circuits output in binary bit strings to the neural network, as shown in Fig. \ref{vqnhe}. The network maps an $n$-bit binary string to a floating point number to yield the following transformation:
\begin{equation}
    |\psi _f\rangle = \sum_{s \in \{0, 1\}^{\otimes n}} f_{\phi}(s) |s\rangle \langle s|\psi\rangle
    \label{neural_transform}
\end{equation}
where $|\psi\rangle$ is the state generated by the ansatz and $f_{\phi}(s)$ denotes the neural network with a set of parameters \textbf{$\phi$}. For simplicity, we write the neural network $f$ without explicitly writing the parameters. The expectation value of the Hamiltonian is
\begin{equation}
\langle \hat{H} \rangle_f = \frac{\langle \psi_f| \hat{H} |\psi_f\rangle}{\langle \psi_f | \psi_f \rangle},
\label{Hamiltonian}
\end{equation}
where it includes the normalization of the expectation value as the transformation given by Eq.~(\ref{neural_transform}) is not unitary.

A method to efficiently evaluate the Pauli terms has also been provided. The qubit corresponding to the first X or Y term in the Pauli term is designated as the star qubit $q^*$. Here on, define $s'\in \{0,1\}^{\otimes n}$ as the bit string with its star qubit set to $0$ while leaving the other qubits, i.e. $s = q_0q_1...q^*...q_n \rightarrow s' = q_0q_1...0...q_n$. Also define the tilde transformation $s \rightarrow \tilde{s}_P$ as follows: for a Pauli letter in the Pauli string $P \in \{I,X,Y,Z\}$, flip the bit if it is either $X$ or $Y$, and leave otherwise. For example, for a bit string $s = 110$ and a Pauli string $P = IXY$, $\tilde{s}_P = 101$. As the Pauli letter corresponding to the star qubit is either $X$ or $Y$, this transformation always flips the star qubit $q^*$. Since in $s'$ the star qubit value is $0$ at all times, the star qubit value in $\tilde{s'}_P$ is always $1$. The expectation value for each Pauli term $\hat{P}$ becomes
\begin{align}
\label{numer}
\begin{split}
    \langle \psi_f| \hat{P} |\psi_f \rangle &= \sum_{s} (-1)^{q^*} f(s') f(\tilde{s'}_P) |\langle s|\psi_m(P)\rangle|^2,
\end{split}
\end{align}
where $|\psi_m(P)\rangle$ represents the ansatz with an additional measurement circuit determined by the Pauli string $P$. The measurement circuit applies controlled-$X$/$Y$ gates on the non-star qubits corresponding to the $X$ or $Y$ strings with the star qubit as the control, followed by basis transformations for measurement on the given Pauli basis. Such transformation yields diagonalization of the Pauli string~\cite{zhang22}, allowing for efficient evaluation of the expectation value with respect to the transformed state. The overall workflow is graphically described in Fig.~\ref{vqnhe}. In the following sections, we call the ansatz circuits with the attached transformations as measurement circuits.

\paragraph*{Divergence of VQNHE for sub-exponential measurement shots.}

Although the algorithm produces a lower estimate of the ground state energy under statevector simulator, VQNHE exhibits critical divergence toward extremely large negative values when executed on actual quantum hardware with limited measurements. This behavior has been discovered through shot-based Qiskit sampler simulations.~\cite{javadi2024}

The neural transformation of the state generated from the ansatz is in general a non-unitary transformation, regardless of the output range of the neural network. To take care of this, VQNHE has included normalization in the final step of evaluating the ground state energy in Eq.~(\ref{Hamiltonian}). However, we have discovered that successful normalization requires an extremely large number of shots, growing exponentially with the number of qubits (See Appendix). This issue arises because the loss function involves bit strings from both the ansatz and measurement circuits, and successful normalization requires that all contributing bit strings be sampled at least once. If any bit string appearing in the numerator of the loss function, originating from the measurement-combined circuits, is absent from the ansatz measurement outcomes, the neural network can exploit the mismatch and induce divergence in the loss. Using the coupon collector analysis, ensuring full coverage demands $O(2^n \log N_M)$ shots on the ansatz circuit, $N_M$ representing the set of bit strings that serve as inputs to the neural network in the numerator.

\begin{figure}

\begin{minipage}[t]{0.88\columnwidth}
\centering
\hspace*{-1.2em}
\includegraphics[width=\textwidth]{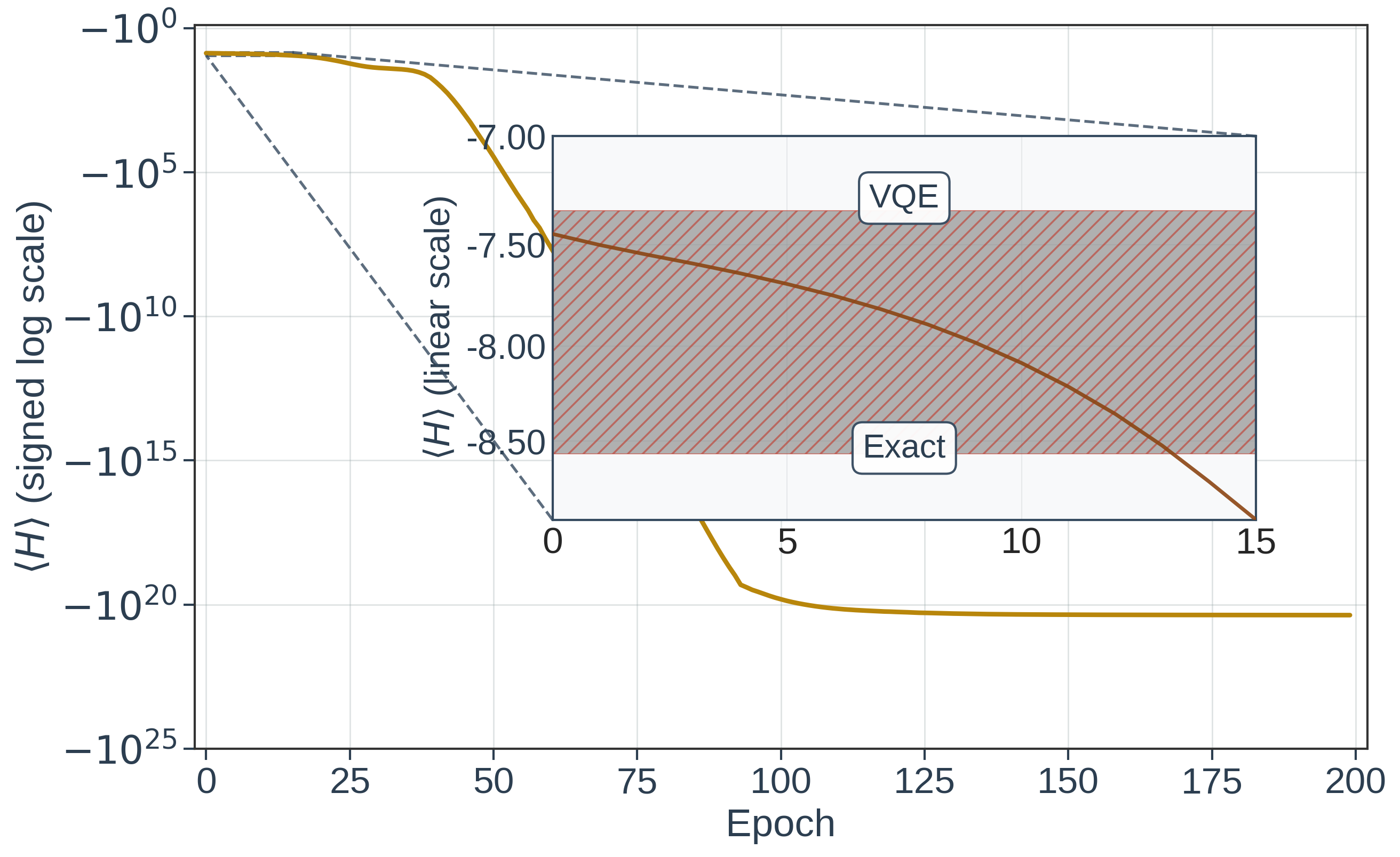}
\textbf{(a)}
\end{minipage}

\vspace{0.5em}

\begin{minipage}[t]{0.82\columnwidth}
\centering
\includegraphics[width=\textwidth]{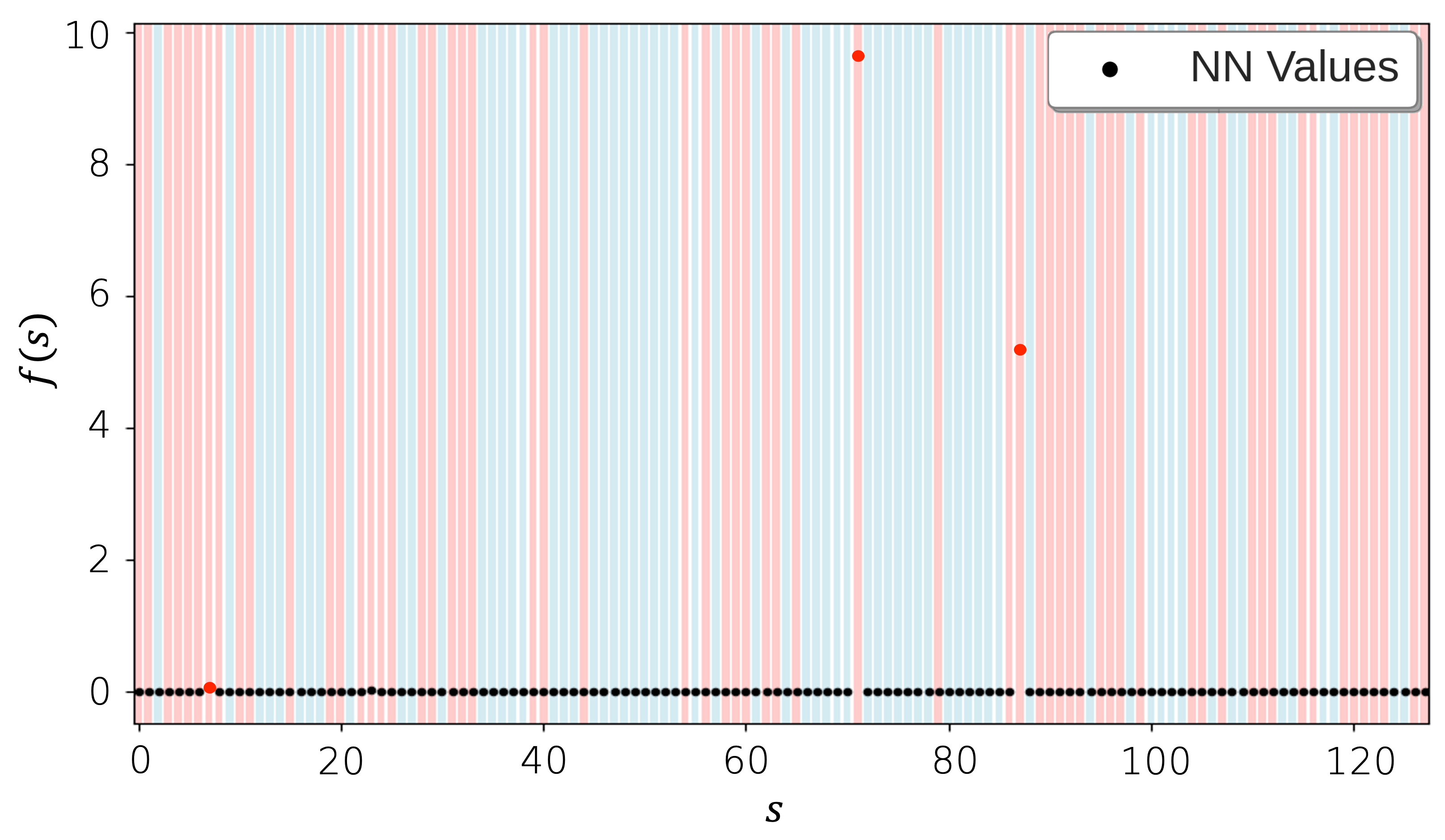}
\textbf{(b)}
\end{minipage}

\caption{
VQNHE implementation of a 7-site TFIM with 7 qubits.
(a) Training of the neural network in VQNHE. The vertical axis shows the loss function—the expectation value of the Hamiltonian—on a log scale. The quantum circuit simulation uses the Qiskit sampler with 500 shots per circuit. The inset highlights the region between the lowest energy from the bare VQE (i.e., VQE without a neural network) and the exact ground state energy; values below this are invalid.
(b) Output of the trained neural network after 200 epochs. The horizontal axis shows bit strings (in decimal), and the vertical axis shows neural network outputs. Blue lines mark bit strings that were sampled (observed) from the ansatz, and red lines mark bit strings that were never observed. Dots mark the network output values. While most values lie near $10^{-6}$ (shown by black dots), extreme values (above $10^{-2}$, shown in red) appear. The neural network assigns extremely large output values to some of those unobserved (red) strings, effectively contributing only to the numerator of the loss (since they are absent in the denominator).
}
\label{fig:7site}
\end{figure}

We present a failure of VQNHE empirically using Qiskit sampler along with the landscape of the trained neural network to support the aforementioned claim in Fig. \ref{fig:7site}. The result quickly falls below the exact ground state energy of the Hamiltonian as shown in Fig. \ref{fig:7site}(a), as it diverges to values lower than $-10^{20}$. The addressed failure has been further confirmed by observation of the values of the neural network after training, which Fig. \ref{fig:7site}(b) displays. The neural network had been trained to yield extreme values for inputs of bit strings that are missing from the ansatz circuit. Thus, if the number of shots is insufficient to produce all $2^n$ bit strings from the ansatz circuit, VQNHE cannot ensure a reliable optimization of the ground state energy.

\paragraph*{Inaccurate behavior for larger shots.}

Until now, it has been argued that the algorithm requires at least an exponential number of shots to prevent any plausible divergence of the ground state estimate. Unfortunately, it turns out that even if such behavior is prevented by increasing the number of shots, VQNHE can still fail to evaluate the ground state energy accurately, falling below it. Looking back at the evaluation of the expectation value of the transformed state of Eq.~(\ref{numer}), each value of the neural network is multiplied by $|\langle s|\psi_m(P)\rangle|^2$, which is the probability of measuring $s$ out of the measurement circuit of the Pauli string $P$. 

With finite number of measurements, there is an unavoidable statistical error between the reconstructed probability and the actual probability. This error is then multiplied by the values of the neural network. Thus, depending on the range of values of the neural network and the number of shots, the result can deviate from the expected result from the theoretical VQNHE execution. The deviation can be statistically modeled by the variance of the difference of the expectation values of Hamiltonian with respect to the exact state $\langle \hat{H} \rangle$ and the state evaluated with finite shots $\langle \hat{H} \rangle_m$. This variance depends on the number of shots $N$ and the values of the neural network (see the Appendix)
\begin{equation}
    \text{Var}[\langle \hat{H} \rangle - \langle \hat{H} \rangle_m] = \frac{\Gamma_f}{N} + \frac{\Delta_f}{N^2},
\end{equation}
where $\Gamma_f$ and $\Delta_f$ are values that depend on the neural network values and the quantum states (See Supplementary Material for detailed derivation~\cite{supp}.) The key lies in the neural network dependence of $\Gamma_f$ and $\Delta_f$. Although the closed-form expression tends to overestimate, it indicates that the neural network seeks to minimize the expectation value once optimization space is available. Therefore, without constraining the output range of the neural network strictly, the deviation can become large enough to significantly distort the result.

\begin{figure}[htb]
    \centering
    \includegraphics[width=0.8\columnwidth]{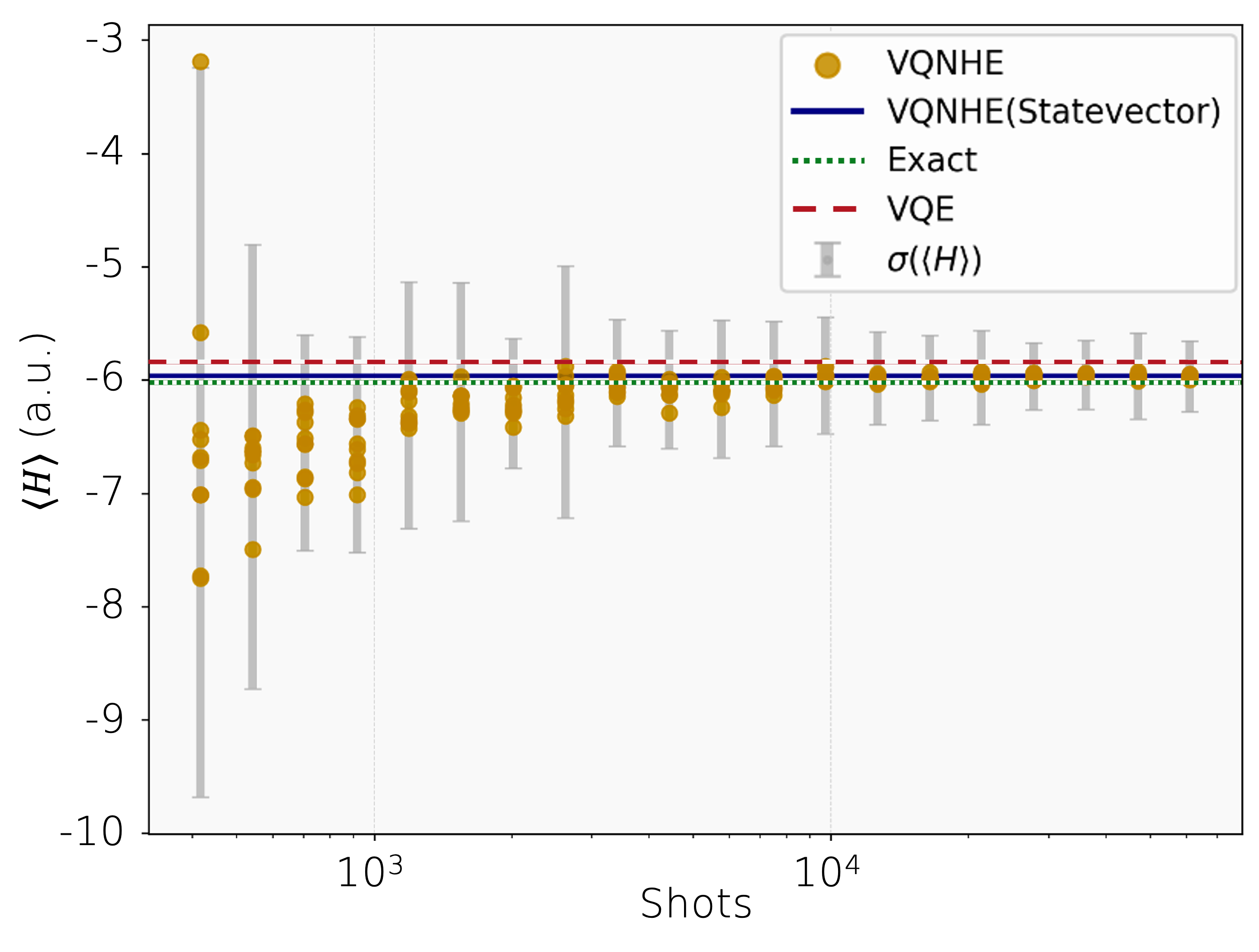}
    \caption{%
    VQNHE results for different numbers of shots, where all bit strings are measured from the ansatz circuit. Yellow markers represent individual VQNHE trials with shots ranging from 500 to 50,000. The parameters of the PQC are fixed throughout. Ideally, values should fall between the two dotted lines, the upper of which represents the exact VQE value and the lower represents exact ground state energy. However, without large number of shots, VQNHE significantly deviates from this appropriate range. Error bars indicate the standard deviation, $\sigma([\langle \hat{H} \rangle - \langle \hat{H} \rangle_m])$, computed using $p_m(s;P)$ and $p_a(s)$ from the exact VQNHE and the neural network values, averaged over trials at each shot count.}
    \label{fig:5q_shots}
\end{figure}

The behavior is clearly demonstrated in Fig. \ref{fig:5q_shots}. Note that quantum circuit measurements without all $2^5=32$ bit strings present were excluded from the sample, as they experience severe divergence as mentioned in the previous section. As the overall variance scales with $\frac{1}{N}$ and $\frac{1}{N^2}$ terms, the VQNHE results converge to the exact VQNHE value (shown in blue line) at larger $N$. However, without sufficient number of shots, it falls below the exact ground state energy, which is not a valid solution. One thing to notice is the existence of poorly trained results lying significantly above the exact VQE results due to the neural network stuck in a wrong minimum due to discrepancies of the measured probabilities and the actual ones. Overall, even with all bit strings measured at least once, which already costs an exponentially large number of measurements, the incongruity between the shot-based VQNHE and its exact version is still large, resulting in even worse approximation to the ground state energy than the VQE alone at times.

\paragraph*{Unitary-VQNHE}

\begin{figure*}[tb]
\centering

\begin{minipage}[t]{0.33\textwidth}
    \centering
    \includegraphics[width=\textwidth]{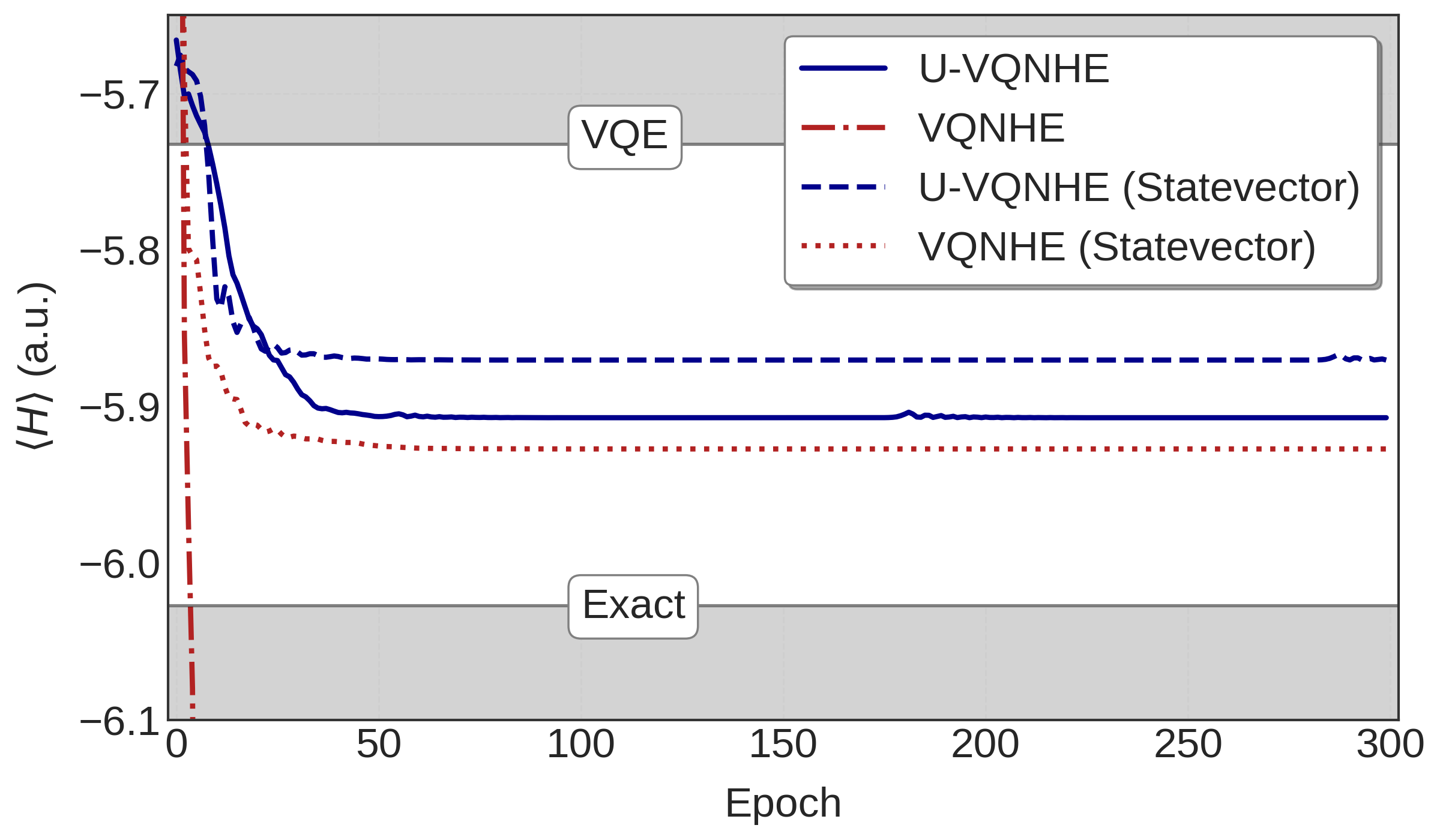}
    \textbf{(a)}
\end{minipage}
\begin{minipage}[t]{0.33\textwidth}
    \centering
    \includegraphics[width=\textwidth]{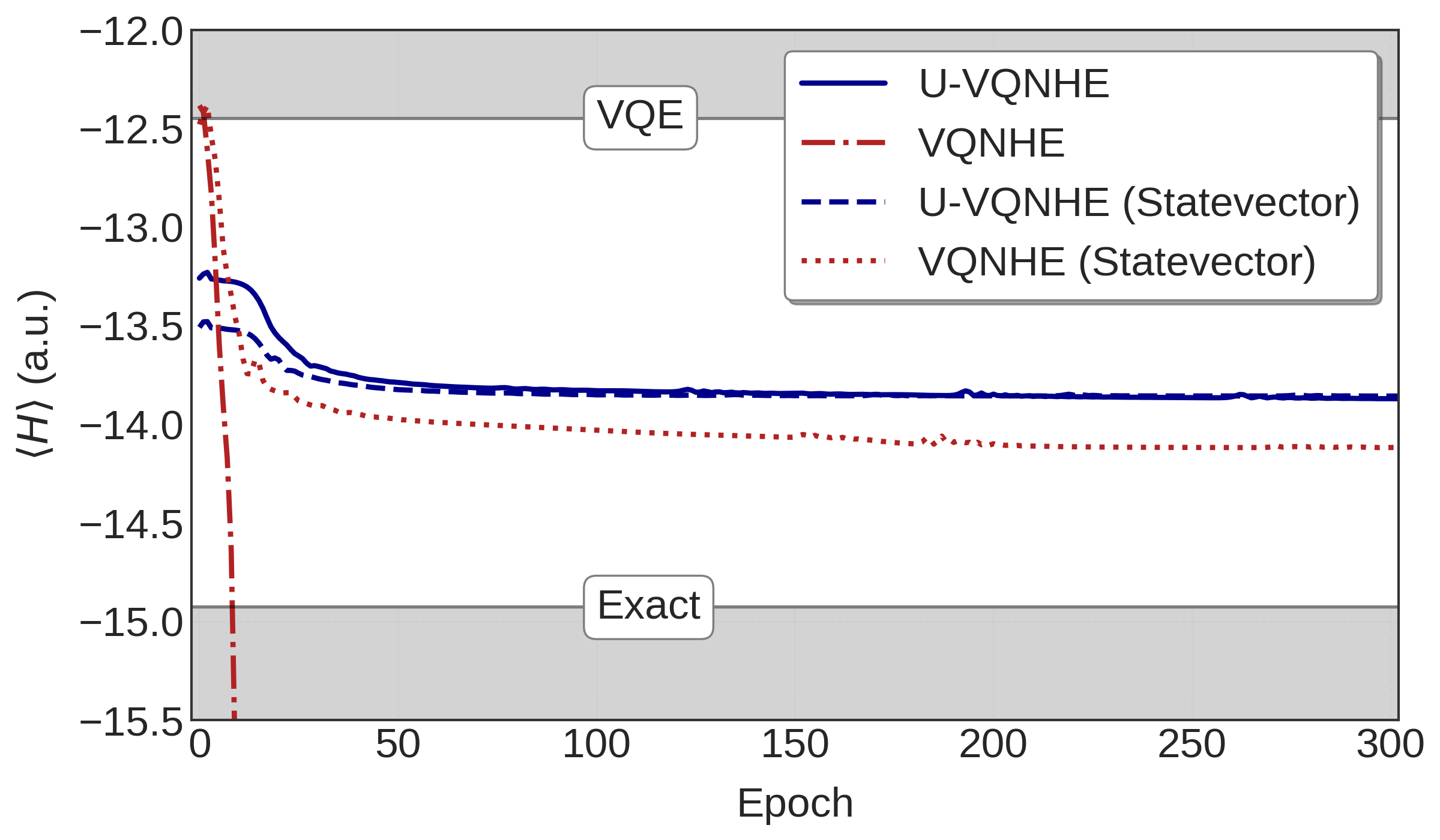}
    \textbf{(b)}
\end{minipage}
\begin{minipage}[t]{0.31\textwidth}
    \centering
    \includegraphics[width=\textwidth]{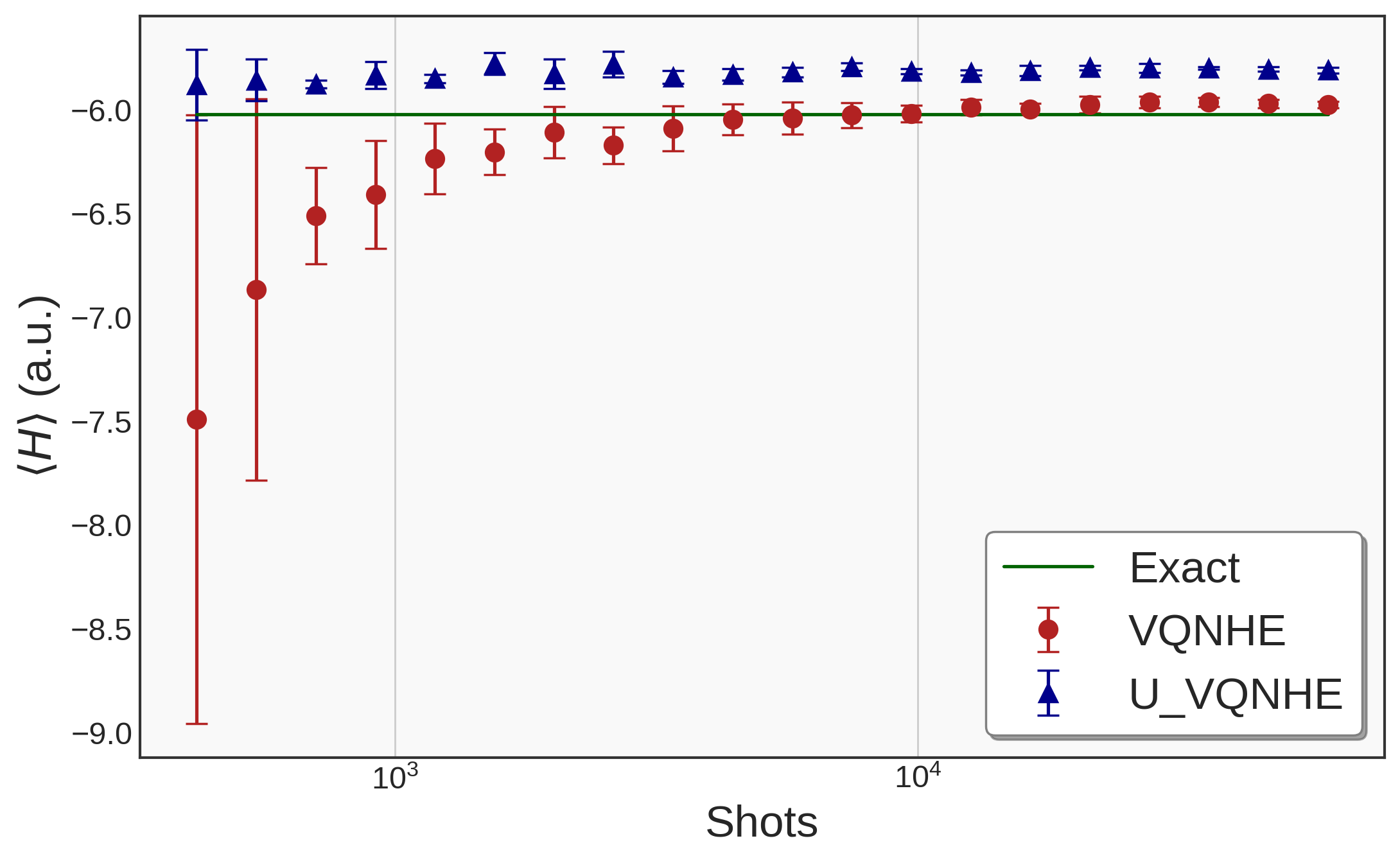}
    \textbf{(c)}
\end{minipage}

\caption{%
Comparison between the training performance of VQNHE and U-VQNHE. 
(a) Result for a 5-site TFIM using a single-layer hardware-efficient ansatz and 100 shots. 
(b) Result for a 12-site TFIM using a two-layer ansatz and 5000 shots. 
In both cases, the region between the exact ground state energy and the VQE result is shown in white, while all other regions are shaded in gray. Solid lines represent results from shot-based simulators, and dashed lines are from statevector simulators. 
(c) U-VQNHE results of 5-site TFIM for different numbers of shots compared to VQNHE. Unlike VQNHE, U-VQNHE does not fall below the exact ground state by a significant amount, showing more stable results even with a small number of shots. The number of shots used was greater than $2^n$ to highlight this stability.}
\label{fig:final}
\end{figure*}

To overcome the limitations of VQNHE, we propose an alternative algorithm that not only requires polynomial scale of computation overhead but is also convergent to a value larger than the ground-state energy regardless of the number of shots. The Unitary-Variational Quantum-Neural Hybrid Eigensolver (U-VQNHE) eliminates the need for normalization, which previously demanded an exponential number of shots for the algorithm to function effectively.

The need for normalization is removed by modifying the way the neural network is applied. Viewing $|\psi\rangle$ as a statevector of $2^n$ elements, the transformation can be expressed as a $2^n \times 2^n$ diagonal matrix. If the transformation is unitary, then the target state $|\psi_f \rangle$ resides in the n-qubit Hilbert space $\mathcal{H}_n$, and the normalization process is no longer needed.

Considering the properties of unitary matrices, in order for a diagonal matrix to be a unitary matrix, each entry $f(s)$ must satisfy $f(s)^\dagger f(s) = 1$. For any real output, the only possibility is that $\forall s, f(s)=\pm1$, under which the neural network cannot represent any significant information. To design a unitary transformation with better expressiveness, the VQNHE must accommodate complex numbers, which has already been formalized (See Supplementary Material of Ref.~\cite{zhang22}). Therefore, we restructured the algorithm to implement the following transformation
\begin{equation}
    |\psi_u\rangle = \sum_{s\in \{0,1\}^{\otimes n}} e^{ig_\phi(s)} |s\rangle \langle s|\psi\rangle,
\label{unitarytransformation}
\end{equation}
in which $g_\phi(s)$ is the outcome of the neural network itself. The neural network structure that evaluates $g_\phi(s)$ follows that of the neural network of VQNHE and can be found in the supplementary materials~\cite{supp}. As long as $g_\phi(s)$ stays real, its specific value does not affect the result, the transformation is always unitary and $|\psi_u\rangle$ is a valid quantum state in $\mathcal{H}_n$. The expectation value from this transformed state $\langle \hat{H} \rangle _g$ becomes
\begin{align}
\langle \hat{H} \rangle_g 
&= \sum_P h_P \big[ \sum_{s \in B_{m,P}}  (-1)^{q^*} \operatorname{Re} \left( e^{-ig(s')} e^{ig(\tilde{s'}_P)} \right) p_m(s;P) \nonumber \\
&+ \sum_{s \in B_{m',P}} (-1)^{q^*} \operatorname{Im} \left( e^{-ig(s')} e^{ig(\tilde{s'}_P)} \right) p_{m'}(s;P) \big].
\label{uvqnhe}
\end{align}

Here, $p_m(s;P)$ is the probability distribution of the measurement circuits of VQNHE, and $p_{m'}(s;P)$ is that of the measurement circuits corresponding to the imaginary parts. For the imaginary part evaluations, one uses Hadamard gate for $X$ and $RX(\pi/2)$ gate for $Y$ letters in the basis transformation stage. The total number of measurement circuits that must be evaluated is at most doubled, but none of them requires exponential number of shots, nor does the algorithm require normalization circuit.

Fig.~\ref{fig:final} compares U-VQNHE with the standard VQNHE, both with statevector simulator and shot-based sampler for the execution of quantum circuits. The dashed lines show the results from the statevector simulator~\cite{javadi2024}. Sampler results with finite shots are shown in solid lines. As previously shown, shot-based VQNHE quickly plunges to large negative values, whereas U-VQNHE stays in the region between the exact ground state energy and the optimized value of VQE even with the sampler. Although U-VQNHE does not achieve expectation values as low as those from VQNHE, it is far more suitable for implementation on shot-based quantum hardware given current hardware limitations.

In addition to relaxing the requirement that all $2^n$ basis bit strings of $\mathcal{H}_n$ be present, thereby avoiding the divergence issues seen in VQNHE, U-VQNHE also exhibits smaller deviations from the exact ground state energy when all bit strings are included. Fig.~\ref{fig:final}(c) shows how the number of shots affects each algorithm. The key factor making U-VQNHE more stable is that the variance of VQNHE shows dependencies on magnitudes of the neural network outputs $f(s)$. Standard VQNHE has no specific restrictions on their values, whereas U-VQNHE restricts their absolute values to $1$. When scaling up the system, the uncertainty due to lack of shots is a huge bottleneck, and U-VQNHE mitigates the uncertainty of VQNHE.

\paragraph*{Discussions and Outlooks}

In this research, it has been shown that while VQNHE possesses a powerful ability to enhance the expressiveness of VQEs with hardware-efficient ansatze, actual implementation of the algorithm requires an exponentially large number of measurements on the quantum circuits in order for the optimization to stay within the bound. Moreover, even if all $2^n$ bit strings expressed with $n$ qubits are present out of the ansatz circuit, which is a condition that must be strictly kept to prevent the divergence, VQNHE failed to produce accurate values for the ground state of $n$-site TFIM due to the inherently large variance. It comes from the fact that the values of the neural network can be large.

To address these issues to apply the neural network in a fully scalable fashion, we have suggested unitary-VQNHE. The main focus is to avoid the necessity to use the normalization at all, which have been the cause of anomalous behavior, while achieving comparable expressiveness. To do so, we have redesigned the neural network to yield a complex value. The suggested algorithm has been shown to yield the ground state energy of 5- and 12-site TFIM stably and limit the variance owing to the unitarity of the neural network outputs. By resolving scalability issues of VQNHE while maintaining its advantages in resource efficiency, U-VQNHE not only enhances stability and applicability of the direct applications of VQNHE~\cite{zhang23, zhang25}, but also possesses potential for quantum advantage in the near future for ground state estimation of complex Hamiltonian with simple ansatz structures.

\paragraph*{Acknowledgments}

This work was supported by the National Research Foundation (NRF) of Korea Grant No. RS-2024-00413957 and No. RS-2024-00442855, funded by the Korean government (MSIT).

\paragraph*{Author Contributions}

M.K. conceived the project, identified the scalability issues, proposed the algorithm, developed the mathematical framework, and wrote the manuscript. K.P. provided guidance on the project and feedback on the manuscript. The simulation code was implemented by M.K., with contributions from U.J. and S.L. T.K. supervised the project, including research planning and manuscript preparation. We also thank Luning Zhao for mentoring and helpful discussions.

\paragraph{Code Availability}

Python scripts that generate all figures and execute the algorithms are archived at Github,
\url{https://github.com/miinukim/vqnhelib}.

\appendix

\section{APPENDIX}

\section{Requirement of exponential measurements in VQNHE}

Training of the neural network utilizes Eq.~(\ref{Hamiltonian}) as its loss function. For a successful implementation of ground state evaluation, the loss function must not fall below the exact ground state energy regardless of the values of $f_\phi$. However, if any bit string from the ansatz circuit’s output is missing, the neural network can exploit this gap because the numerator might include a term absent from the denominator. Let $B_{a}$ be the set of bit strings that have been measured at least once upon measurement of the ansatz circuit and $B_{m, P}$ as that of the measurement-combined circuits corresponding to a Pauli term $P$, and $p_a(s)$ and $p_m(s;P)$ be the the probability distributions of bit strings with respect to the ansatz and the combined circuits:
\begin{equation}
p_a(s)=|\langle s|\psi\rangle|^2, p_m(s;P)=|\langle s|\psi_m(P)\rangle|^2.
\end{equation}
One can then rewrite the loss function in terms of the neural network results as
\begin{equation}
    L=\frac{\sum_P h_P \sum_{s \in B_{m, P}}{(-1)^{q^*} f(s') f(\tilde{s'}_P)p_m(s;P)}}{\sum_{s \in B_a} f(s)^2 p_a(s)}
\label{loss}
\end{equation}

Denote the entire set of bit strings that the numerator consists of as $B_M = \bigcup_P [\{s'|s \in B_{m,P}\} \cup \{\tilde{s'}_P|s \in B_{m,P}\}]$. If $\exists s^* \in B_M \backslash B_a,$ then $f(s^*)$ in the numerator of Eq.~(\ref{loss}) is unmatched in the denominator, meaning that it is outside the scope of normalization and therefore can take a very large value, inducing divergence of the loss function. The argument also applies even if one constrains the domain of the output of the neural network to positive values by, for example, applying a softmax activation at the final layer (See Supplementary Material for detailed explanation~\cite{supp}.)

The minimum number of shots required to guarantee that $B_M \subset B_a$ remains to be determined. Suppose $B_M$ has $N_M$ elements, which can have up to $2^n$ of them. Evaluating the average number of shots that must be executed in the ansatz circuit in order to include every element of $B_M$ in the most general case is a tricky problem, but its lower bound can be obtained in the case of equal probability. If the ansatz circuit yields every bit string with equal probability, the expected number of required shots for the ansatz $N_a$ is
\begin{equation}
    \mathbb{E}[N_a]=2^n H_{N_M}
\end{equation}
according to the coupon collector's problem~\cite{neal08}, where $H_{N_M} = \sum_{i=1}^{N_M} \frac{1}{i} \approx \log(N_M)$. This means that for successful implementation of VQNHE, the number of shots $N_a \in O(2^n \log {N_M})$ is required on the ansatz circuit for successful training. For general quantum states, this number is typically larger because an unbalanced distribution makes it much harder to sample outcomes with low probabilities. For a quantum algorithm to assert its scalability, it must not require exponential resource for successful implementation. Consequently, VQNHE does not scale efficiently because its shot requirements grow over exponential scale.

\section{Theoretical inaccuracy of VQNHE}

Assume that each circuit has been measured $N$ times. Let $c_a(s)$ be the count of incidents where the bit string $s$ is measured from the ansatz circuit and $c_m(s;P)$ from the measurement circuit corresponding to the Pauli string $P$. The reconstructed probability distribution of the ansatz and measurement circuits are $p_{a,R}(s)=\frac{c_a(s)}{N}$ and $p_{m,R}(s;P)=\frac{c_m(s;P)}{N}$. Expectation value of the Pauli string $P$ from finite measurements $\langle P \rangle_m$ is estimated using $p_{a,R}(s)$ and $p_{m,R}(s;P)$ rather than the true distributions $p_a(s)$ and $p_m(s;P)$. Assuming noiseless implementation of quantum circuits, $c_a(s)$ and $c_m(s;P)$ follow binomial distributions
\begin{align}
    c_a(s) &\sim B(N, |\langle s | \psi \rangle |^2) \notag \\
    c_m(s;P) &\sim B(N, |\langle s | \psi_m(P) \rangle |^2).
\end{align}
The variance of $\langle P \rangle - \langle P \rangle_m$ can be evaluated by assuming that the covariances of $c_a(s)$ and $c_a(s')$ is 0 if $s \neq s'$, given that there are exponentially many bit strings such that each value does not significantly disturb others. (See Supplementary Material for detailed derivation~\cite{supp}.)

In addition, when evaluating the variance of the expectation value of the Hamiltonian $\hat{H}=\sum_P h_P \hat{P}$, one must take the covariances between the expectation values of each of the Pauli terms into account. The overall variance of the error in expectation value of the Hamiltonian from VQNHE is 
\begin{equation}
\begin{split}
    \text{Var}[\langle \hat{H} \rangle - \langle \hat{H} \rangle_m] &= \sum_P |h_P|^2 \text{Var}[\langle P \rangle - \langle P \rangle_m] 
    \\ & + \sum_{P\neq P'} h_P h_{P'}\text{Cov}[\langle P_1\rangle_m, \langle P_2\rangle_m].
    \\ & = \frac{{\Gamma_f}}{N} + \frac{{\Delta_f}}{N^2},
\label{var}
\end{split}
\end{equation}
(Detailed derivation can be found in Supplementary Material~\cite{supp}). Specifically, they depend on the values of the neural networks $f(s)$ and the exact probability distributions $p_a(s)$ and $p_m(s;P)$ of the quantum states. Since one cannot exactly predict the training process of the neural network as it is done after completing the measurements of the quantum circuits, the calculated variance describes the trend over $N$ regarding extent to which the shot-based VQNHE deviates from the statevector version of it.

\bibliographystyle{apsrev4-2}
\bibliography{references}

\clearpage
\onecolumngrid
\appendix
\section*{Supplementary Materials}
\input{supplementary}

\end{document}

%% file: supplementary.tex
\section{Divergence of VQNHE with positive neural network values}

The loss function in the main text is the target of the neural network to minimize. 

\begin{equation}
    L=\frac{\sum_P h_P \sum_{s \in B_{m, P}}{(-1)^{q^*} f(s') f(\tilde{s'}_P)p_m(s;P)}}{\sum_{s \in B_a} f(s)^2 p_a(s)}
\label{loss}
\end{equation}

When the values of the neural network $f(s)$ are strictly limited to positive values by applying a softmax activation, for example, the divergence seems less obvious. In this section we explain how the loss function can still be trained to a large negative value under positive-valued neural networks.

At the baseline, there must exist an element in $B_M\backslash B_a$, which we denote as $s^*$. This is the divergence condition for general neural network. Now, viewing in terms of $f(s^*)$, in the Loss function each of it is multiplied by
\begin{equation}
    \sum_P \sum_{q^* \in {0,1}} h_P(-1)^{q^*}f(\tilde{s^*_P})p_m(s;P),
\end{equation}
where in this case, the star qubit can either take $0$ or $1$. For elements in $B_M \backslash B_a$, the condition for divergence under positive-valued neural network is the following: given a Hamiltonian and probability distribution, there exists at least one element $s^{**}$ such that $h_P(-1)^{q^*}f(\tilde{s^{**}_P})p_m(s;P) < 0$ for some neural network $f(s)$. In other words, if the neural network can manage to find a way to yield $f(\tilde{s^{**}_P)}$ values such that this value is negative, positively growing $f(s^{**})$ results in negative increase of the loss function, causing the divergence.

\section{Derivation of the variance of VQNHE}

In this section, derivation of $\text{Var}[\langle H \rangle - \langle H \rangle_m]$ in the main text is explained in detail. As mentioned in the main text, we denote the exact probability amplitudes of the bit strings given by the quantum state generated by the ansatz circuit as $p_a(s)=|\langle s|\psi\rangle|^2$, and that of the measurement circuit corresponding to a Pauli string $P$ as $p_m(s;P)(s)=|\langle s|\psi_P\rangle|$. From repetitive measurements of the quantum circuits, one can only yield $p_{a,R}(s)=\frac{c_a(s)}{N}$ and $p_{m,R}(s;P)=\frac{c_m(s;P)}{N}$ with inevitable inaccuracy, $N$ being the number of measurements.

Assume there exists no error source so that each and every measurement is ideal. Given a state $|\psi\rangle$, probability of measuring $s\in \{0,1\}^{\otimes n}$ is $p_a(s)$, and is independent of the values of the previous measurements, letting us assert that $c_a(s)\sim B(N, |\langle s | \psi \rangle|^2), c_m(s;P)\sim B(N, |\langle s | \psi_P \rangle|^2)$. In addition, given large enough number of qubits, there are exponentially many possible bit strings as an outcome, each affecting a small portion viewing as a whole. We thus assume that all $c_a(s)$ are independent probability variables, i.e. $\text{Cov}(c_a(s), c_a(s'))=0, \forall P, \text{Cov}(c_m(s;P), c_m(s';P))=0$ if $s \neq s'$. Now, writing the expectation value of each Pauli term P out of VQNHE as $\langle P \rangle$ out of exact probability distribution and $\langle P\rangle_m$ out of measured probabilities,

\begin{align}
    \langle P \rangle = \frac{\sum_s (-1)^{q^*}f(s')f(\tilde{s'}_P)p_a(s)}{\sum_s f(s)^2 p_a(s)} , \langle P \rangle_m = \frac{\sum_s (-1)^{q^*}f(s')f(\tilde{s'}_P)p_a(s)_m}{\sum_s f(s)^2 p_{m,R}(s;P)}
\end{align}

Here, we introduce, for simplicity, the probability distribution errors $\epsilon(s)=p_a(s)-p_{a,R}(s), \epsilon_P(s)=p_m(s;P)(s)-p_{m,R}(s;P)$. Then,

\begin{equation}
    \sum_s (-1)^{q^*} f(s')f(\tilde{s'}_P)p_m(s;P)(s) = \langle P \rangle \sum_s p_a(s)f(s)^2 = \langle P \rangle\sum_s (p_{a,R}(s)f(s)^2 - \epsilon(s)f(s)^2)
\end{equation}

\begin{equation}
    \sum_s (-1)^{q^*} f(s')f(\tilde{s'}_P)p_{m,R}(s;P) = \langle P \rangle_m \sum_s p_{a,R}(s)f(s)^2
\end{equation}

Thus,

\begin{equation}
    \langle P \rangle - \langle P \rangle _m= \frac{\sum_s (-1)^{q^*}f(s')f(\tilde{s'}_P)\epsilon_P(s)}{\sum_s f(s)^2 p_{a,R}(s)}
    - \frac{\sum_s f(s)^2 \epsilon(s)}{\sum_s f(s)^2 p_{a,R}(s)} \langle P \rangle
\label{p-pm}
\end{equation}

The task now on is to evaluate the variance of $\langle P \rangle - \langle P\rangle _m$ in the case of taking $N$ measurements out of each circuit. Note that $p_a(s)$ and $p_{a,R}(s)$ follow binomial distributions and thus each have variance

\begin{gather}
    \text{Var}[p_{a,R}(s)] = \text{Var}\left[\frac{c_a(s)}{N}\right] = \frac{p_a(s)(1-p_a(s))}{N}, \text{Var}[p_{m, P}(s)] = \frac{p_m(s;P)(1-p_m(s;P))}{N}
\end{gather}

From this, we evaluate the variance of the difference between $\langle P \rangle$ and $\langle P \rangle_m$.

\begin{equation}
\begin{split}
    &\text{Var}\Bigg[\sum_s f(s') f(\tilde{s'}_P) \epsilon_P(s) 
    - \sum_s f(s)^2 \epsilon(s) \langle P \rangle \Bigg] \notag \\
    &= \frac{1}{N} \sum_s \Big[f(s') f(\tilde{s'}_P) 
    \big\{p_m(s;P)(s) (1 - p_m(s;P)(s))\big\} 
    + f(s)^2 |\langle P \rangle| p_a(s)(1 - p_a(s))\Big],
\end{split}
\end{equation}

since $\epsilon(s)$ and $\epsilon_P(s)$ are independent. Evaluating the numerator, applying Taylor expansion about the average,

\begin{equation}
\begin{split}
    &\text{Var}\Big[ \frac{1}{\sum_s f(s)^2 p_{a,R}(s)} \Big] = \frac{1}{[\sum_s f(s)^2 p_a(s)]^4} 
    \cdot \frac{1}{N} 
    \sum_s f(s)^2 p_a(s)(1 - p_a(s)).
\end{split}
\
\end{equation}

Since

\begin{equation}
\begin{split}
    \text{Var}\left[\frac{A}{B}\right] &\approx 
    \text{Var}[A] \, \text{Var}\left[\frac{1}{B}\right] 
    + \text{Var}[A]\left(\mathbb{E}\left[\frac{1}{B}\right]\right)^2 
    + \text{Var}\left[\frac{1}{B}\right]\left(\mathbb{E}[A]\right)^2
\end{split}
\end{equation}

and as for some coefficient $C, C'$, $\mathbb{E}[A] = C \mathbb{E}[\epsilon_P(s)] + C' \mathbb{E}[\epsilon(s)] = 0$, the variance of the error becomes

\begin{equation}
\begin{split}
    \text{Var}[\langle P \rangle - \langle P \rangle_m] = 
    & \Bigg\{ \frac{1}{N^2} \frac{\sum_{s} f(s)^2 p_a(s)(1-p_a(s)) }{[\sum_s f(s)^2 p_a(s)]^4} 
    + \frac{1}{N} \frac{1}{[\sum_s f(s)^2 p_a(s)]^2} \Bigg\} \\
    & \times \sum_s \Bigg\{f(s')f(\tilde{s'}_P) 
    \Big[p_m(s;P)(s) (1-p_m(s;P)(s))\Big] 
    + f(s)^2|\langle P \rangle| p_a(s) (1-p_a(s))\Bigg\}.
\label{variance}
\end{split}
\end{equation}

In addition, unlike the covariances between different $c_a(s)$ and $c_m(s;P)$ terms, which can be considered to be independent, when evaluating $\text{Var}[\langle H \rangle - \langle H \rangle_m] = \sum_P |h_P|^2 \text{Var}[\langle P \rangle - \langle P \rangle_m] + \sum_{P\neq P'} h_P h_{P'}\text{Cov}[\langle P_1\rangle_m, \langle P_2\rangle_m]$, the covariances require close attention, as evaluation of each Pauli term shares the common ansatz for normalization. With the notations introduced before, one can write covariance as

\begin{equation}
\begin{split}
    \text{Cov}[\langle P_1\rangle_m, \langle P_2\rangle_m] = \text{Cov} \Bigg[ \sum_s \frac{(-1)^{q^*} f(s')f(\tilde{s'_{P_1}})p_{P_1}(s)}{\sum_s f(s)^2p_a(s)}, \frac{(-1)^{q^*} f(s')f(\tilde{s'_{P_2}})p_{P_2}(s)}{\sum_s f(s)^2p_a(s)}\bigg] = Cov\bigg[\frac{A}{C}, \frac{B}{C}\bigg]
\end{split}
\label{cov1}
\end{equation}
where $p_{P_1}(s)$ and $p_{P_2}(s)$ are independent, as they come from different measurement circuits. For simplicity, we assume that each Pauli term uses different circuit, whereas the number of measurement circuits can be reduced for different Pauli strings with same $X$ and $Y$ letters. In such cases, all terms involving the random variables $p_m(s;P)(s)$ and $p_a(s)$ are equivalent upon sign difference. Accordingly, those terms can be merged into a single term with its sign factor modified to $(-1)^{q^*}h_P + (-1)^{q^*}h_{P'}$.

Evaluating Eq.~(\ref{cov1}) is straightforward, as they only share the denominator. For some $A, B, C$, using second-order Taylor expansion around the mean of $C$,

\begin{equation}
\begin{split}
    \text{Cov}\bigg[\frac{A}{C}, \frac{B}{C}\bigg] = \mathbb{E}\bigg[\frac{AB}{C^2}\bigg] - \mathbb{E}\bigg[\frac{A}{C}\bigg]\mathbb{E}\bigg[\frac{B}{C}\bigg] \approx \frac{\mathbb{E}[A]\mathbb{E}[B]\text{Var}[C]}{\mathbb{E}[C]^4}
\end{split}
\end{equation}

Thus, applying to Eq.~(\ref{cov1}),

\begin{equation}
\begin{split}
    \text{Cov}[\langle P_1\rangle_m, \langle P_2\rangle_m] = \frac{[\sum_s (-1)^{q^*}f(s')f(\tilde{s'_{P_1}})p_{P_1}(s)][\sum_s (-1)^{q^*}f(s')f(\tilde{s'_{P_2}})p_{P_2}(s)][\sum_s f(s)^2 p_a(s)(1-p_a(s))]}{N [\sum_s f(s)^2 p_a(s)]^4}
\label{cov2}
\end{split}
\end{equation}

Evaluating Eq.~(\ref{variance}) and Eq.~(\ref{cov2}) for the Pauli terms composing the Hamiltonian, one can evaluate the approximate variance of $\langle H \rangle - \langle H \rangle_m$. 

\begin{equation}
    \text{Var}[\langle H \rangle - \langle H \rangle_m] = \frac{\Gamma_f}{N} + \frac{\Delta_f}{N^2},
\end{equation}

\begin{equation}
\begin{split}
    \Xi &= \sum_s \Bigg\{f(s')f(\tilde{s'}_P) 
    \Big[p_m(s;P)(s) (1-p_m(s;P)(s))\Big] 
    + f(s)^2|\langle P \rangle| p_a(s) (1-p_a(s))\Bigg\} \\ 
    \\ \Gamma_f &= \sum_P |h_P|^2 \Bigg\{ \frac{1}{[\sum_{s^*} f(s^*)^2 p_a(s^*)]^2} \Bigg\}\Xi  \\
    &+ \sum_{P\neq P'} \frac{h_P h_{P'}[\sum_s (-1)^{q^*}f(s')f(\tilde{s'_{P_1}})p_{P_1}(s)][\sum_s (-1)^{q^*}f(s')f(\tilde{s'_{P_2}})p_{P_2}(s)][\sum_s f(s)^2 p_a(s)(1-p_a(s))]}{[\sum_s f(s)^2 p_a(s)]^4} \\
    \Delta_f &= \sum_{P} |h_P|^2 \frac{\sum_{s} f(s)^2 p_a(s)(1-p_a(s)) }{[\sum_s f(s)^2 p_a(s)]^4}\Xi.
\end{split}
\end{equation}

Note that this is a slight overestimation of the variance, since $\mathbb{E}[1/X]>1/\mathbb{E}[X]$. As the entire variance is mainly proportional to $1/N$ with additional $1/N^2$ terms due to the variance of the denominator, the entire variance disappears at infinite shots, as expected. In addition, in overall, all terms show heavy dependence on the values of $f(s)$. Similar to what has been asserted in the authors of Ref.~\cite{zhang22}, wide distribution of the values $f(s)$ is a critical reason for large deviation of VQNHE. On the other hand, although the exact formula does not directly apply, U-VQNHE heavily constrains the values of output of the neural network, reducing the deviation in general.

\section{Possibilities of measurement-derived divergence for VQE variants with non-unitary transformations.}

VQNHE is not the only algorithm that utilizes non-unitary transformation onto VQE. Several research efforts attempt to apply additional gates and post-processing in a non-unitary fashion in order to achieve better performance in terms of accuracy, especially for the eigensolvers with hardware-efficient ansatze. We briefly review potentials for measurement-derived divergence discussed in our work for several stuides that utilize non-unitary transformations onto the states and discuss their resource efficiencies compared to our work.

One such research is called the non-unitary variational quantum eigensolver(nu-VQE)\cite{benfenati2021}. It performs non-unitary operation as a set of quantum gates applied on top of the prepared ansatz. In specific, the work makes use of the single-body and two-body Jastrow factors\cite{jastrow1955} as its non-unitary process. The Jastrow operator takes the form
\begin{equation}
\begin{split}
    &J =J_1 + J_2, \\
    &J_1 = \text{exp}\Big[ -\sum_{i=1}^{N} \alpha_i Z_i \Big], J_2 = \text{exp}\Big[ -\sum_{i<j}^{N} \lambda_{i,j} Z_i Z_j \Big],
\label{jastrow}
\end{split}
\end{equation}
where $\alpha_i$ and $\lambda_{i,j}$ are trainable parameters, equivalent to the parameters of the neural network in the case of VQNHE. The overall Jastrow operator is approximated to its linear form for simplicity
\begin{equation}
    J(\vec{\alpha}, \vec{\lambda})=1 -\sum_{i=1}^{N} \alpha_i Z_i -\sum_{i<j}^{N} \lambda_{i,j} Z_i Z_j.
\label{jastrowlinear}
\end{equation}
As acting $J(\vec{\alpha}, \vec{\lambda})$ is non-unitary, similar to VQNHE, nu-VQE requires normalization to obtain the energy.
\begin{equation}
\begin{split}
    E &= \frac{\langle \psi(\vec{\theta}) | \hat{J}^\dagger (\vec{\alpha}, \vec{\lambda}) \hat{H} \hat{J} (\vec{\alpha}, \vec{\lambda}) | \psi(\vec{\theta}) \rangle}{\langle \psi(\vec{\theta}) | \hat{J}^\dagger (\vec{\alpha}, \vec{\lambda}) \hat{J} (\vec{\alpha}, \vec{\lambda}) | \psi(\vec{\theta}) \rangle} \\
    &= \sum_{\hat{P}} c_P 
    \frac{
        \langle \psi | \hat{P} | \psi \rangle 
        - \sum_i \alpha_i \langle \psi | \hat{P} \hat{Z}_i | \psi \rangle 
        - \sum_{i<j} \lambda_{i,j} \langle \psi | \hat{P} \hat{Z}_i \hat{Z}_j | \psi \rangle
    }{
        \langle \psi | \psi \rangle 
        - \sum_i \alpha_i \langle \psi | \hat{Z}_i | \psi\rangle 
        - \sum_{i<j} \lambda_{i,j} \langle\psi|\hat{Z}_i\hat{Z}_j|\psi\rangle
    }
\end{split}
\label{jastrowexp}
\end{equation}
The equation on the second line of Eq.~(\ref{jastrowexp}) is obtained by writing down the Jastrow factor(Eq.~(\ref{jastrowlinear}) explicitly in its linear form.

Going through the scenarios of divergence in the paper, the first obvious possibility is that viewing the parameters $\vec{\alpha}, \vec{\lambda}$ as variables, any coefficient on the denominator yields zero and the corresponding one on the numerator gives nonzero value. This corresponds to the case where bit strings out of measurement process are missing on the denominator side, causing severe failure of the algorithm. Nevertheless, this is not a realistic situation in terms of nu-VQE. In order to satisfy the conditions, it must be the case where $\exists i$ such that $\langle\psi | \hat{Z_i} | \psi\rangle =0 , \langle\psi | \hat{P} \hat{Z_i} | \psi\rangle \neq 0$ or $\exists i, j$ such that $\langle\psi | \hat{Z_i} \hat{Z_j} | \psi\rangle =0 , \langle\psi | \hat{P}\hat{Z_i}\hat{Z_j} | \psi\rangle \neq 0$. Although not theoretically impossible, getting an exact zero as an expectation value is highly unlikely, unless one happens to prepare an exactly balanced state as $|\psi\rangle$ or the number of measurement outcomes yielding –1 for $Z_i$ equals the number yielding +1. In such rare cases, it might experience similar failure to VQNHE, although this is not considered a major flaw of the algorithm.

Despite its idea of applying a non-unitary transformation without significant risk of divergence and thus without concerns of scalability bottleneck due to number of shots, nu-VQE requires many more quantum circuit evaluations than VQNHE or U-VQNHE. As written in Eq.~(\ref{jastrowlinear}), in order to provide the expectation values to all the terms, it requires at most $O(n^4)$ quantum circuit evaluations for each Pauli term comprising the target Hamiltonian. Note that this is minimal overhead of using the Jastrow factor at its linear approximation. Any higher-ordered terms require much more circuits, and the exact Jastrow factor demands exponential of them. Thus, VQNHE and U-VQNHE require much less computational resources in terms of the quantum computations, while requiring only polynomially scaling classical resources.

Another attempt to apply non-unitary transformation, named Jastrow quantum circuit (JQC)\cite{mazzola2019}, applies the transformation variationally on the ansatz state itself while mapping operators to the qubit space via, for example, Jordan-Wigner transformation\cite{reiner2016}. The projector that implements Jastrow operator is written as
\begin{equation}
    \mathcal{P}_J = e^J, J = \sum_{k\neq l} \lambda_{kl}Z_k Z_l.
\end{equation}
Note that it captures the second-order terms of Eq.~(\ref{jastrow}) while not exploiting its linear approximation form. In Ref.~\cite{mazzola2019} it is argued that the nonunitary Jastrow operator effectively filters out components that yield high energy eigenvalues.

Although the detailed implementation and approximation details differ, JQC and nu-VQE actually end up with the same procedure from the expectation value point-of-view. The difference comes from the target on which the Jastrow operator acts. Thus, in terms of possible divergence, the same arguments on nu-VQE apply for non-unitary transformations by projective mapping of states. In addition, in terms of resource efficiency, it experiences the same overhead in the number of quantum circuits required, making VQNHE and U-VQNHE superior considering the resource overhead.

Alternatively, there exists a variant of VQE that exploits non-unitary transformation in a different fashion. Cascaded variational quantum eigensolver (CVQE)\cite{gunlycke2024} constructs its ansatz in the Fock space, which is mapped to the qubit space similar to the JQC, and applies the following additional non-unitary transformation in the occupation number eigenstates
\begin{equation}
    \hat{\lambda}(\phi)=\sum_{n \in \mathcal{N}} \lambda_n(\phi) |n\rangle \langle n|,
\end{equation}
which lets the transformed state to be written as
\begin{equation}
    |\psi(\phi)\rangle =e^{i\hat{\lambda}(\phi)}\hat{U}|0\rangle = e^{i\hat{\lambda}(\phi)}|\psi_0\rangle,
\end{equation}
where $U$ is a unitary operator and $|0\rangle$ represents the vacuum state, written in the second quantization picture. It can be a fixed operator or be trained with simple ansatze, just as hardware-efficient ansatz block is trained in VQNHE before optimizing parameters of the neural network. The result of the expectation value of the given Hamiltonian can then be written as
\begin{equation}
    E = \frac{
    \langle\psi_0|e^{i\hat{\lambda^\dagger}(\phi)}\hat{H}e^{i\hat{\lambda}(\phi)}|\psi_0\rangle
    }{
    \langle\psi_0|e^{i\hat{\lambda^\dagger}(\phi)}e^{i\hat{\lambda}(\phi)}|\psi_0\rangle
    }.
\end{equation}

The key difference between the previous algorithms and CVQE is that each variational parameter $\lambda_n(\phi)$ is associated with a number state, or a Fock state $|n\rangle$, whereas for the previous algorithms the parameters were associated with the Jastrow terms. As there are exponentially many terms comprising $e^{i\hat{\lambda^\dagger}(\phi)}\hat{H}e^{i\hat{\lambda}(\phi)}$, CVQE expresses the expectation values as that linear combination of operators $\{\hat{R}_{lm}\}$, which can be obtained analytically, that express the expectation value in a diagonal form. Note that the scheme is similar to VQNHE. In fact, CVQE can be regarded as a generalization of VQNHE to a complex domain and second quantization picture. Due to the direct dependence of the parameters, the gradient of the energy with respect to the parameters $\phi$ can be analytically obtained. Also note that the given transformation on the ansatz state resembles that of U-VQNHE, except that U-VQNHE restricts the values of $\lambda_n$ to real numbers.

So is CVQE vulnerable to divergence during optimization? Again, an important factor is whether, in the expectation value evaluation, any parameter in the denominator is multiplied by zero. The expectation value, written in terms of the set of operators $\{\hat{R^\dagger}_{lm}\}$, is
\begin{equation}
\begin{split}
    & \Gamma(\phi)=\sum_{l\in \mathcal{L}} \sum_{m \in \mathcal{M}_l} \sum_{n \in \mathcal{N}} v_{lmn} e^{-i\lambda^*_{\dot{n}_l^+\vec{n}}(\phi)} e^{i\lambda_{\dot{n}_l^-\vec{n}}(\phi)} |\langle \psi_0|\hat{R^\dagger}_{lm}|n\rangle|^2 \\
    & \Lambda(\phi) = \sum_{n\in \mathcal{N}} e^{-2 \text{Im}\lambda_n(\phi)} |\langle\psi_0|n\rangle|^2 \\
    & E = \frac{\Gamma(\phi)}{\Lambda(\phi)},
\end{split}
\end{equation}
where $\mathcal{L}$ represents indices of the terms in Hamiltonian expressed as $\hat{H} = \sum_{l \in \mathcal{L}} h_l C^\dagger_{n_l^+} C_{n^-_l}$, $\mathcal{M}_l$ represents the basis of the Hilbert space spanned by the supports of each term $C^\dagger_{n_l^+} C_{n^-_l}$, and $v_{lmn}$ is a coefficient depending on these indices. Note, similar to how VQNHE is implemented on a shot-basis, that $|\langle \psi_0|\hat{R^\dagger}_{lm}|n\rangle|^2$ and $|\langle \psi_0|n\rangle|^2$ are the counts of each bit strings(Jordan-Wigner transformation\cite{reiner2016} maps each number state $|n\rangle$ into qubit state $\bigotimes_{q\in \mathcal{Q}} |n_q\rangle \in \mathcal{H}_Q$). In addition, each $\lambda_n(\phi)$ is multiplied by $|\langle \psi_0|\hat{R^\dagger}_{lm}|n\rangle|^2$ on the numerator and $|\langle \psi_0|n\rangle|^2$ on the denominator. Thus, it follows that CVQE is very likely to diverge, unless it is given with an exponential scale of shots so that all elements in $\mathcal{H}_Q$ are measured at least once from $|\psi_0\rangle$.

Analogous to our work, in order to maintain the structure of CVQE while preventing aforementioned divergence under reasonable amount of shots, it must be guaranteed that $\forall n, \lambda_n(\phi) \in \mathbb{R}$, which resembles the unitary transformation from the neural network of the main text. By confining the parametrized operator $\hat{\lambda}(\phi)$ thus to be unitary, CVQE loses its ability to seek beyond the Fock space as asserted by its authors, although the actual solution lives in the Fock space. However, combining how CVQE handles diagonalization of the terms comprising the Hamiltonian described in the Fock space onto the framework of U-VQNHE presents potentials for a resource-efficient variational algorithm applicable in the second quantization formalism. The lack of expressiveness of the postprocessing alone can be leveraged by utilizing a hardware-efficient VQE as a unitary operator acting on the vacuum state and train both parameters in sequence. In addition, with use of hardware-efficient ansatz for the initial unitary, the post-processing can adopt neural network for generality and expressiveness.

In short, there are several other variants of VQE that utilizes non-unitary transformation to enhance the expressiveness. While algorithms that makes use of Jastrow factor to modify the quantum state mapping\cite{mazzola2019} or the Hamiltonian\cite{benfenati2021} are unlikely to experience the divergence during the optimization of the parameters of the non-unitary transformations, they require much more quantum resources than VQNHE or U-VQNHE and thus are not as resource-efficient. On the other hand, CVQE\cite{gunlycke2024} displays a lot of resemblance to VQNHE in terms of how they evaluate the expectation value with respect to the parameters, and our analysis indicates that CVQE also requires an exponential amount of shots to be taken to evaluate the denominator to prevent itself from divergence issues. Thus, no algorithm among those listed above presents better resource-efficiency than U-VQNHE considering both the number of quantum circuits to be evaluated and the number of shots to be taken.

\section{Notes on training of the parameters}

For studies that involve training of neural networks, setting the hyperparameters related to its training is of significant importance. However, in terms of the neural network used in this research, which consists of two fully-connected layers and a series of activation functions, including ReLU\cite{bai2022}, sigmoid and tanh functions, training of it is not as difficult of a work compared to state-of-the-art neural network researches. Nevertheless, there are several aspects of it worth mentioning.

The VQNHE and U-VQNHE system consist of two sets of parameters: variational parameters for the quantum circuits and the parameters for the classical neural network. All simulations in the research have been conducted by training the VQE first, and then training the neural network on top of the already trained VQE. The reason for that is that, possible as it may seem, joint training of the parameters is impossible in the case of VQNHE, or in general, quantum algorithms with neural post-processing. When the VQE output is fixed, the neural network evaluation is deterministic, and its optimization is stable, whereas when the output of the quantum circuit is not fixed, it yields deviations on every measurement process. For joint training, this is a critical limitation, as the VQE yields unstable outputs and the neural network struggles to find a way to be trained such that it consistently provides downward gradient. Moreover, repetitive evaluation of the quantum circuit is very costly, making joint training very inefficient.

In terms of optimization tools, gradient-free COBYLA optimizer~\cite{Powell94} has been used for optimizing the parameters of VQE and ADAM for the neural network. Nevertheless, one can also choose to use parameter-shift rule~\cite{wierichs22} to obtain the quantum gradients of the given ansatz circuit and use gradient-based optimization techniques, such as stochastic gradient-descent optimization. In order to apply the parameter-shift rule, all the parametrized gates must have their generators with two unique eigenvalues. As our choice of hardware-efficient ansatz consists only of RX and RZZ gates as parametrized gates, they all fall under the criterion, allowing for easy usage of the parameter-shift rule.